\documentclass{article}
\usepackage[preprint]{spconf}

\usepackage{graphicx} 
\usepackage[utf8]{inputenc}
\usepackage[T1]{fontenc}
\usepackage[hidelinks]{hyperref}
\usepackage{amsmath,amssymb,amsfonts}
\usepackage{siunitx}
\usepackage{booktabs}
\usepackage{xcolor}
\usepackage{soul}
\usepackage{svg}

\title{Physics-Informed Anomaly Detection of Terrain Material Change in Radar Imagery}

\name{\parbox{\linewidth}{\centering Abdel Hakiem Mohamed Abbas Mohamed Ahmed$^{1,2}$, Beth Jelfs$^{1}$, Airlie Chapman$^{2}$, Eric Schoof$^{3}$, Christopher~Gilliam$^{1}$}}
\address{$^1$School of Engineering, University of Birmingham, UK\\ $^2$Department of Mechanical Engineering, University of Melbourne, Australia\\ $^3$Department of Electrical and Electronic Engineering, University of Melbourne, Australia\\  Email: axm1614@student.bham.ac.uk, \{b.jelfs, c.gilliam.1\}@bham.ac.uk,\\ \{airlie.chapman, eschoof\}@unimelb.edu.au}

 \copyrightnotice{\copyright\ IEEE 2026 Personal use of this material is permitted. Permission from IEEE must be obtained for all other uses, in any current or future media.}

\begin{document}
\ninept
\maketitle
\begin{abstract}
    In this paper we consider physics-informed detection of terrain material change in radar imagery (e.g., shifts in permittivity, roughness or moisture). We propose a lightweight electromagnetic (EM) forward model to simulate bi-temporal single-look complex (SLC) images from labelled material maps. On these data, we derive physics-aware feature stacks that include interferometric coherence, and evaluate unsupervised detectors: Reed-Xiaoli (RX)/Local-RX with robust scatter (Tyler's M-estimator), Coherent Change Detection (CCD), and a compact convolutional auto-encoder. Monte Carlo experiments sweep dielectric/roughness/moisture changes, number of looks and clutter regimes (gamma vs K-family) at fixed probability of false alarm. Results on synthetic but physically grounded scenes show that coherence and robust covariance markedly improve anomaly detection of material changes; a simple score-level fusion achieves the best F1 in heavy-tailed clutter. 
\end{abstract}


\section{Introduction}
Detecting subtle, spatially localised changes in terrain materials from radar imagery underpins applications in infrastructure monitoring, environmental surveillance and security~\cite{Rosen2000, UlabyLong2014, ElDarymli2013}. In coherent radar, long-standing strategies for detecting changes span intensity-based methods and coherent change detection (CCD), which exploits interferometric coherence estimated between phase-registered image pairs with material or structural changes reducing the coherence magnitude~\cite{BamlerHartl1998,Touzi1999,Wahl2016}. While state-of-the-art deep learning models have advanced bi-temporal change mapping, most works target pixel stacks statistically, with limited explicit ties to the EM material properties that actually derive radar backscatter and decorrelation~\cite{KhelifiMignotte2020,Bai2023}.

The radar backscatter obtained from terrain depends on a number of factors, such as the complex permittivity and surface roughness parameters of the terrain, as well as the radar geometry (incidence angle). Widely used rough-surface models such as the Integral Equation Model (IEM/AIEM) provide validated relationships between these parameters and the radar backscatter~\cite{UlabyLong2014,FungChen2004}, whereas, popular anomaly detectors like the Reed-Xiaoli (RX) score assume a (often Gaussian) background distribution and flag departures from this via Mahalanobis distance~\cite{ReedYu1990}. In radar imagery, heavy-tailed clutter is common, so a fixed global threshold does not preserve a stable false alarm rate across heterogenous backgrounds; using robust covariance estimators (e.g., Tyler's M-estimator) makes detectors more CFAR-like, i.e., the decision threshold adapts to local clutter statistics so the false alarm probability remains approximately constant across the scene~\cite{ElDarymli2013,Tyler1987}.

Despite the different approaches to anomaly detection in radar imagery, there remains a lack of reproducible, controlled studies that link material-level changes (e.g., moisture-driven permittivity or roughness changes) to expected changes in backscatter and coherence, and the relative performance of RX/robust-RX, CCD and modern unsupervised models. Material change (e.g., dry\(\to\)wet soil, asphalt ageing, vegetation removal) alters permittivity and roughness, shifting Fresnel/rough-surface reflectivity and speckle statistics; simultaneously, coherence decreases in changed regions. Thus, coherence-aware features and robust background modelling are well-matched not only to radiometric change but also to material change~\cite{UlabyLong2014,ElDarymli2013,Touzi1999}.   

In this work, we address the problem of terrain material change by proposing a physics-informed anomaly detection framework. We use a compact, IEM-inspired forward model to map labelled material maps to bi-temporal single look complex (SLC) images. Per-pixel parameters for permittivity, rms height, correlation length and incidence angle determine mean backscatter; multiplicative speckle and controlled cross-epoch correlation generate SLC pairs that emulate decorrelation for genuine material/structural change~\cite{UlabyLong2014,FungChen2004}. From these images, a physics-aware feature stack is estimated in local windows combining \(\log\)-intensities, simple texture, incidence angle, and interferometric coherence \(\hat{\gamma}\)~\cite{Touzi1999}. We then perform a comparative study of unsupervised detectors to highlight regimes where physics-aware features and robust covariance are decisive. We consider both global and local RX with a robust scatter estimator for heavy-tailed clutter~\cite{ElDarymli2013,ReedYu1990,Tyler1987}; CCD via decorrelation~\cite{Wahl2016}; and a lightweight convolutional auto-encoder trained on unchanged tiles~\cite{KhelifiMignotte2020,Daudt2018}.

The remainder of this paper is structured as follows: Section~\ref{sec:sim} details the proposed forward model and Section~\ref{sec:method} describes the features and detectors. In  Section~\ref{sec:exp} we outline the datasets used and the Monte-Carlo protocol which covers changes in permittivity, rms height, number of looks and Signal to Noise Ratio (SNR). Section~\ref{sec:results} presents results and ablation data reporting Receiver Operating Characteristic-Area Under the Curve (ROC-AUC), Average Precision (AP) and F1 score at low Probability of False Alarm (PFA). Section~\ref{sec:conclusion} concludes the paper, highlighting that on synthetic, but physically grounded data, coherence-aware methods decisively outperform traditional intensity-based detectors, with a simple score-level fusion achieving the highest performance.

\section{Physics-Informed Forward Model}\label{sec:sim}

In this section, we describe our SLC image formation framework using a physics-informed forward model. In our model we assume a monostatic, side-looking geometry with a known radar pose. A base digital elevation model (DEM) provides local surface normal $\mathbf{n}(\mathbf{r})$ (with $\mathbf{r}$ the geospatial location in the DEM/frame); the local incidence angle is $\theta(\mathbf{r})=\arccos(\mathbf{\hat{k}}\!\cdot\!\mathbf{n})$, where $\mathbf{\hat{k}}$ is a unit vector denoting the look direction. A labelled material map assigns, per pixel $p$, complex permittivity $\varepsilon_{r,p}(f)$, where $f$ is the radar operating frequency and roughness parameters: rms height $\sigma_p$ and correlation length $l_{c,p}$. These parameters control microwave backscatter according to rough-surface EM models such as IEM/AIEM~\cite{UlabyLong2014,FungChen2004,AIEMUpdate2008,AIEMValidation2015}.

The exact IEM/AIEM backscatter requires spectral integrals and regime checks~\cite{FungChen2004}. Therefore to allow large-scale Monte-Carlo studies, we use a lightweight surrogate that preserves the main dependencies~\cite{Oh1992,Dubois1995}: 
\begin{equation}
    \sigma^{0}_{p}(\theta) \;\approx\;F\!\big(\varepsilon_{r,p},\theta\big)\;\,G\!\big(\theta,\sigma_p,l_{c_p},f\big),
    \label{eq:IEMlite}
\end{equation}
where $F$ is a Fresnel term and $G$ is a roughness attenuation inspired by IEM/AIEM. For a single polarisation (e.g., vertical co-planar, VV), we take:
\begin{equation}
    F(\varepsilon_r,\theta) = 
\left|\frac{\varepsilon_r\cos\theta - \sqrt{\varepsilon_r-\sin^2\theta}}{\varepsilon_r\cos\theta + \sqrt{\varepsilon_r-\sin^2\theta}}\right|^2, 
\end{equation}
and a smooth attenuation such as 
\begin{equation}
G(\theta,\sigma,l_c,f)\;=\;\exp\!\big[-(2k\sigma\cos\theta)^2\big]\;\phi(l_c),\quad k=2\pi f/c,
\end{equation}
with $\phi(l_c)$ a bounded, monotone factor capturing correlation-length effects. Although~\eqref{eq:IEMlite} is not a substitute for AIEM, it tracks the trend that higher $\Re\{\varepsilon_r\}$ (or moisture) and appropriate roughness alter $\sigma^{0}$ systematically~\cite{UlabyLong2014,FungChen2004}. Where fidelity is crucial, we swap $G$ for the closed-form IEM terms within a restricted validity regime~\cite{AIEMUpdate2008,AIEMValidation2015}.

\subsection{SLC image formation}
Under fully developed speckle, a SLC pixel is modelled as
\begin{equation}
S^{(t)}_{p} \;=\; \sqrt{\sigma^{0,(t)}_{p}}\,\eta^{(t)}_{p}\,e^{j\phi^{(t)}_{p}},
\end{equation}
where $\eta$ is circular complex Gaussian (unit mean intensity) and $\phi$ includes deterministic phase from geometry. $L$-look intensities follow a Gamma distribution; heterogeneous terrain often exhibits heavy-tailed, compound-Gaussian (e.g., $K$) statistics, impacting CFAR behaviour~\cite{ElDarymli2013}.

Between two epochs $(t_1,t_2)$, we synthesise correlated speckle via:
\begin{equation}
W_2 \;=\; \gamma\,W_1 + \sqrt{1-|\gamma|^2}\,W_\perp,\qquad W_1,W_\perp\sim\mathcal{CN}(0,1),
\label{eq:correlatedspeckle}
\end{equation}
using a spatially varying coherence parameter $\gamma(\mathbf{r})$ and where $\mathcal{CN}$ is the complex normal distribution. In unchanged areas $|\gamma|\!\approx\!1$ (after coregistration), whilst material/structural changes reduce $|\gamma|$~\cite{Touzi1999}. The sample coherence over a window $\mathcal{W}$ is:
\begin{equation}
\hat{\gamma}=\frac{\sum_{p\in\mathcal{W}} S^{(t_1)}_p \,{S^{(t_2)}_p}^{*}}{\sqrt{\sum_{p\in\mathcal{W}} \left|S^{(t_1)}_p\right|^2}\,\sqrt{\sum_{p\in\mathcal{W}} \left|S^{(t_2)}_p\right|^2}},
\label{eq:coh}
\end{equation}
with bias/variance properties detailed in~\cite{Touzi1999}. For a CCD baseline, we also consider the maximum-likelihood change statistic of~\cite{Wahl2016}, which offers a CFAR-like behaviour with respect to clutter-to-noise ratio. 

Material change is injected by modifying $\{\varepsilon_r,\sigma,l_c\}$ within selected regions (e.g., dry$\to$wet soil, asphalt$\to$gravel). The simulator outputs paired SLCs, intensities, and auxiliary maps (incidence, slope). Limitations (discussed in Section~\ref{sec:conclusion}) include the single-frequency, single-pol default and omission of volume scattering and layover; nonetheless, the setup suffices to study how material perturbations propagate to intensity/coherence statistics~\cite{UlabyLong2014,FungChen2004}.

\section{Features and detectors}\label{sec:method}
We now detail the features and detectors we will use to study the anomaly detection problem.
\subsection{Physics-aware feature stack}
To obtain a representative feature stack, given SLCs $\left(S^{t_1}, S^{t_2}\right)$ we form:
\begin{align}
&I_1 = \big|S^{t-1}\big|^2,\quad 
I_2 = \big|S^{t_2}\big|^2,\quad
R_{\log}=\log\!\frac{I_2+\epsilon}
{I_1+\epsilon},\nonumber\\
&\text{texture}: \{\mu,\sigma^2\}\ \text{over
} N\!\times\!N,\quad \theta\ \text{(incidence)},\quad
|\hat{\gamma}|\ \text{from \eqref{eq:coh}}.
\end{align}
These features jointly encode radiometry, local heterogeneity, geometry and coherence, all of which respond to $\Delta\varepsilon_r$ and roughness changes~\cite{UlabyLong2014,Touzi1999}.

\begin{figure*}[htb]
    \centering
    \includegraphics[width=0.8\linewidth, trim=0 20 0 00, clip]{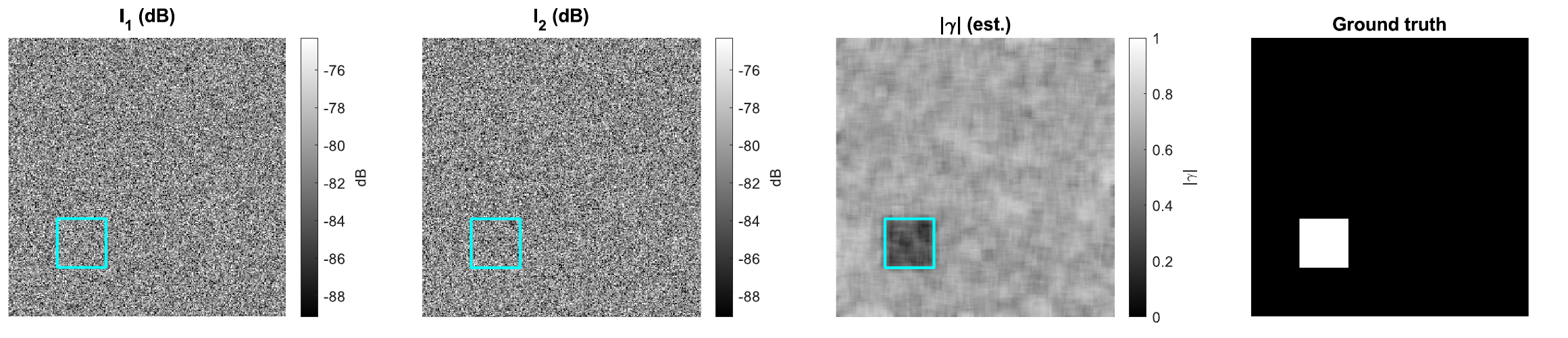}\vspace*{-3mm}
    \caption{Representative trial - feature maps. Left$\rightarrow$right: log-$I_1$, log-$I_2$, coherence magnitude $|\hat{\gamma}|$, ground truth (GT).\vspace*{-5mm}}
    \label{fig:qual}
    \vspace*{-2mm}
\end{figure*}

\begin{figure*}[htb]
    \centering
    \includegraphics[width=0.7\linewidth, trim=0 10 0 0, clip]{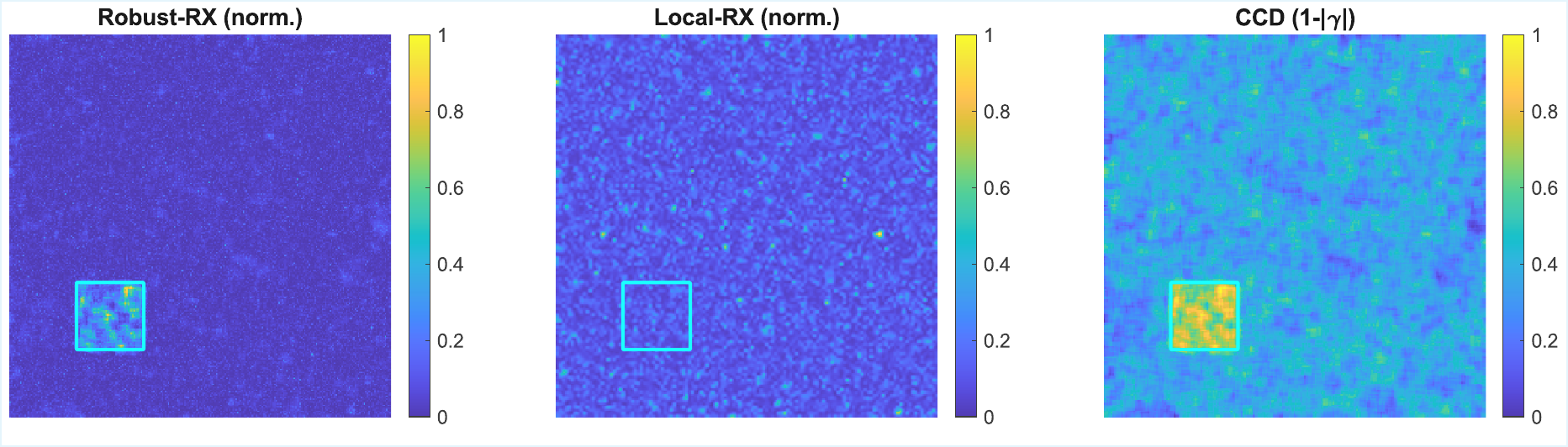}\vspace*{-3mm}
    \caption{Representative trial - detector score maps. Left$\rightarrow$right: RXrob, Local-RX, CCD ($1-|\hat{\gamma}|$). \vspace*{-5mm}}
    \label{fig:detectors}
\end{figure*}

\subsection{Detectors}
In all of our analysis we standardise individual scores from each of the detectors and compute a fused score by a weighted sum or a logistic regressor fitted on a small validation subset. Operating points are selected by fixing a global PFA (e.g., $10^{-3}$) using background pixels. Isolated false alarms are removed via morphological opening/closing. Reporting follows standard detection practice: ROC-AUC, AP and F1 at fixed PFA~\cite{ElDarymli2013}.

\paragraph*{RX and Local-RX anomaly detection}
Stacking features $x\in\mathbb{R}^{d}$, the RX score is the Mahalanobis distance:
\begin{equation}
    D^2(x) = (x-\mu)^\top\Sigma^{-1}(x-\mu),
    \label{eq:rx}
\end{equation}
with $(\mu,\Sigma)$ estimated from a background set; Local-RX uses a sliding neighbourhood to adapt to non-stationary environments~\cite{ReedYu1990}. Under a Gaussian background and known $\Sigma$, $D^2\sim\chi^2_D$, enabling CFAR thresholding; in practice, estimation error and non-Gaussian clutter require robustification.

\paragraph*{Robust scatter for heavy-tailed clutter}
Radar clutter is frequently heavy-tailed (Gamma /$K-$distributed intensities), violating Gaussian assumptions~\cite{ElDarymli2013}. We therefore replace the sample covariance in \eqref{eq:rx} by Tyler's M-estimator~\cite{Tyler1987}, which is distribution-free within the complex elliptically symmetric family: 
\begin{equation}
\Sigma \leftarrow \frac{d}{n}\sum_{i=1}^{n}\frac{(x_i-\mu)(x_i-\mu)^\top}{(x_i-\mu)^\top\Sigma^{-1}(x_i-\mu)},
\quad \text{with }\mathrm{tr}(\Sigma)=d,
\label{eq:tyler}
\end{equation}
where $d$ is the number of features in $x$ and $\mathrm{tr}(\cdot)$ is the trace operator. In our experiments, we use a regularised form: 30 fixed-point iterations with 5\% ridge shrinkage towards a scaled identity for small-sample conditioning. This choice is standard for heavy-tailed radar clutter and helps maintain CFAR-like behaviour.

\paragraph*{Coherent Change Detection (CCD)}
CCD treats change as decorrelation. The simplest score is:
\begin{equation}
    \mathrm{CCD}(p)=1-|\hat{\gamma}(p)|,
\end{equation}
thresholded to control PFA. We also report the maximum-likelihood CCD statistics of~\cite{Wahl2016}, which exploits finite-look statistics and exhibits CFAR properties in speckle.

\paragraph*{Unsupervised Auto-Encoder (AE)}
A compact convolutional AE operates on feature tiles (e.g., $32{\times}32{\times}d$). Trained only on unchanged tiles (from $t_1$ or masked stable regions), it minimises reconstruction loss $\| \hat{X}-X\|_2^2$ and yields an anomaly score $e=\|\hat{X}-X\|$ at test time. Surveys document strong change detection performance from Siamese/AE families, while noting domain shift and limited physical interpretability~\cite{Bai2023,Daudt2018,Jiang2022}.

\section{Simulations}\label{sec:exp}
\begin{table}[tb]
\centering
\caption{Global parameters used in the simulations.}
\label{tab:params}
\setlength{\tabcolsep}{11pt}
\begin{tabular}{ll}
\toprule 
Parameter & Range \\
\midrule
Complex SNR & $12$-$28$\,dB \\
Compound texture shape & $\nu\!\in\![0.3,1.2]$ \\
Residual co-registration jitter & $\sigma_{xy}\!\in\![0.08,0.25]$\,px\\
Residual phase jitter & $\sigma_\phi\!\in\![0.10,0.30]$\,rad\\
Vegetation/volume fraction & $0.08$-$0.18$ \\
Unchanged coherence magnitude & $|\gamma_{\mathrm{bg}}|\!\in\![0.90,0.95]$\\
Changed coherence magnitude & $|\gamma_{\mathrm{chg}}|\!\in\![0.45,0.65]$ \\
Looks & $L\!\in\!\{2,\ldots,8\}$\\
\bottomrule
\end{tabular}
\vspace*{-5mm}
\end{table}
\subsection{Simulation setup}

We run $N{=}200$ Monte Carlo trials on $256{\times}256$ scenes. Each trial samples global nuisance parameters uniformly within physically credible ranges to emulate heterogeneous terrain and acquisition imperfections, see Table~\ref{tab:params}. The compound texture shape allows interpolation between heavy-tailed $K$-family and multilook-Gamma clutter and unchanged coherence has an additional decorrelation factor $0.75$-$0.90$ to represent natural variation. Incidence varies smoothly around $35^\circ$ to reflect a side-looking geometry. These factors follow established clutter/CFAR consideration in SAR target detection under non-Gaussian statistics~\cite{ElDarymli2013}.

Material change is injected by perturbing per-pixel $(\varepsilon_r,\sigma,l_c)$ within a compact region (e.g., square/rectangle/ellipse). Mean backscatter $\sigma^0(\theta)$ is generated by an IEM-inspired surrogate that preserves the Fresnel and roughness/geometry dependencies; where appropriate,  closed-form IEM/AIEM components are used within their validity regimes~\cite{UlabyLong2014,FungChen2004,AIEMUpdate2008,AIEMValidation2015}. This ties simulated radiometry to dielectric/roughness changes, such as dry\(\to\)wet soil or asphalt ageing~\cite{UlabyLong2014}.

Paired SLCs $\left(S^{(t_1)},S^{(t_2)}\right)$ are formed with multiplicative speckle and optional compound texture. Cross-epoch speckle is correlated using \eqref{eq:correlatedspeckle} to provide spatially varying coherence fields; sample coherence $\hat{\gamma}$ is estimated in a $7{\times}7$ boxcar, following classical bias/variance analysis~\cite{Touzi1999}. In addition to $1{-}|\hat\gamma|$, we include the maximum-likelihood CCD statistics of~\cite{Wahl2016} as a likelihood-based comparator. 

\subsection{Detectors, training and fusion}
We evaluate: global RX~\cite{ReedYu1990}; robust-RX using Tyler's $M$-estimator with light shrinkage for heavy-tailed background~\cite{Tyler1987,Pascal2008,Ollila2012}; Local-RX with a dual ring (outer/guard windows $21/9)$; CCD ($1{-}|\hat\gamma|$)~\cite{Wahl2016}; and a compact convolutional autoencoder (AE) trained on unchanged tiles (patch $16$ 50 epochs), reflecting unsupervised/deep change-detection practice~\cite{KhelifiMignotte2020,Bai2023,Daudt2018}. Scores are $z$-normalised and fused either equally or with two/three-way weights learned on a $10\%$ calibration subset per trial.

For a representative qualitative figure, we select the single trial with the highest visibility (coherence-dominated separation) and render feature/score maps with the ground truth contour, together with ROC/PR curves. The looks-dependence of coherence variance explains the strong effect of $L$ on visual separability~\cite{Touzi1999}.
\begin{table*}[htb]
\centering
\caption{Aggregate detection performance (mean $\pm$ 95\% CI over $N{=}200$ trials). }
\label{tab:main_results}
\setlength{\tabcolsep}{11pt}
\begin{tabular}{lccc}
\toprule
\textbf{Method} & \textbf{ROC-AUC} & \textbf{AP (PR-AUC} &\textbf{F1}\\
\midrule
RX (global)             & $0.775\, [0.761,\,0.788]$ & --- & ---- \\
RXrob (Tyler)           & $0.775\, [0.760,\,0.788]$ & $0.157\,[0.139,\,0.180]$ &
$0.086\,[0.071,\,0.106]$\\
Local-RX                & $0.489\,[0.486,\,0.493]$ & $0.019\,[0.018,\,0.020]$ &
$0.001\,[0.001,\,0.002]$ \\
CCD                     & $0.901\,[0.891,\,0.910]$ & $0.401\,[0.371,\,0.432]$ &
$0.304\,[0.276,\,0.332]$ \\
FUSE (equal 2-way)      & $0.891\,[0.880,\,0.899]$ & --- & --- \\
FUSEw (learned 2-way)   & $0.901\,[0.891,\,0.910]$ & $0.399\,[0.368,\,0.429]$  & $0.302\,[0.275,\,0.332]$ \\
FUSE3w (learned 3-way)    & $0.901\,[0.891,\,0.910]$ & $0.390\,[0.359,\,0.420]$  & $0.289\,[0.263,\,0.317]$ \\
AE (compact)              & $0.645\,[0.633,\,0.657]$ & $0.054\,[0.048,\,0.062]$  & $0.014\,[0.012,\,0.018]$ \\
\bottomrule
\end{tabular}
\vspace*{-5mm}
\end{table*}

\begin{figure}[htb]
    \centering
    \includegraphics[width=0.6\linewidth, trim=0 5 0 0, clip]{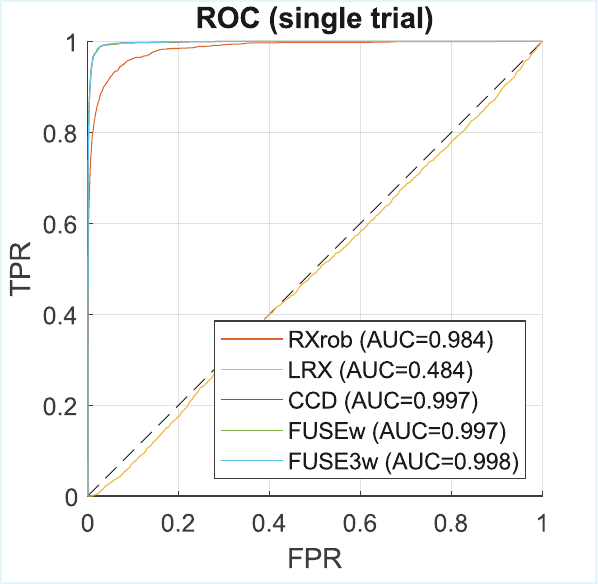}\vspace*{-3mm}
    \caption{Representative trial. ROC computed on the same trials as Figs.~\ref{fig:qual} \ref{fig:detectors}\vspace*{-5mm} }
    \label{fig:ROC}
\end{figure}

\section{Results and Discussion}\label{sec:results}
The results for a representative trial are shown in Figs.~\ref{fig:qual},~\ref{fig:detectors}, and~\ref{fig:ROC}. Figure~\ref{fig:qual} illustrates the log-intensity at $t_1$ and $t_2$, the coherence magnitude and the ground-truth mask. Figure~\ref{fig:detectors} shows the corresponding score maps for RXrob, Local-RX, and CCD. Finally, Fig.~\ref{fig:ROC} shows the ROC results for the trial. These results highlight that the change region is more identifiable in detectors that rely on interferometric coherence. It can be seen that the changed region is more prominent in the interferometric coherence $|\hat{\gamma}|$ in Fig.~\ref{fig:qual} compared to the log-intensity images and this directly translates to the output of the detectors; the region is easily seen in the CCD output in Fig~\ref{fig:detectors} whereas it is very subdued in the Local-RX output. This observation is to be expected as under stable geometry, material/structural change chiefly reduces interferometric coherence $|\hat{\gamma}|$, which CCD targets directly, whereas modest radiometric contrast can be masked by heavy-tailed clutter, and Local-RX’s ring estimates can be contaminated in heterogeneous neighbourhoods. 



Table~\ref{tab:main_results} summarises the aggregate performance across the Monte Carlo simulations. Coherence-centric detectors (CCD and simple fusion) consistently lead in ROC-AUC and on the class imbalance-aware metric (AP and F1 at PFA $= 10^{-3}$). In contrast, intensity-led baselines (RX/Local-RX) do not perform well, with robust scatter giving RX some stability but not closing the gap when decorrelation is the principal cue. These trends agree with the variance properties of the coherence estimator~\cite{Touzi1999}.

In more detail, CCD and coherence weighted fusion score best due to the injected material/structural changes mainly reducing coherence magnitude $|\gamma|$ (decorrelation) whilst radiometric contrast is comparatively modest and partly masked by compound Gaussian clutter. Robust covariance helps RX primarily through CFAR-like stabilisation rather than separability gains. This matches InSAR theory on decorrelation sources and coherence estimator behaviour~\cite{Rosen2000,BamlerHartl1998, Touzi1999} and the need for robust statistics in heavy-tailed radar backgrounds~\cite{ElDarymli2013,Tyler1987,Pascal2008,Ollila2012}. Finally, the low performance of Local-RX is due to it assuming locally stationary background statistics. With vegetation and spatially varying coherence, the dual ring configuration used in Local-RX often mixes populations (changed vs unchanged), contaminating the local mean/covariance and destabilising the computation of inverses for a small window, which explains the weak separation in Table~\ref{tab:main_results}. Similar sensitivity to window design and heterogeneity are reported in anomaly detection studies~\cite{Zhao2015,KwonNasrabadi2005,ElDarymli2013}.

On average our lightweight score fusion is equivalent to CCD. In coherence-dominated regimes, adding RX/AE contributes a limited complementary signal and can actually add variance. Fusion becomes more beneficial when both radiometric and coherent cues carry information (e.g., larger $\Delta\sigma^0$ at higher equivalent number of looks). Consistent with the bias/variance trade off of $\hat\gamma$~\cite{Touzi1999} and CCD practice~\cite{Wahl2016}. 

In terms of sensitivity to the simulation parameters the following trends were observed:
\begin{itemize}
    \item \textbf{Looks ($L$):} higher $L$ lowers the variance $\hat\gamma$, improving CCD and fused metrics~\cite{Touzi1999}. 
    \item \textbf{Texture shape ($\nu$):} heavier tails degrade RX/local-RX more than CCD, motivating robust scatter~\cite{ElDarymli2013,Tyler1987,Pascal2008,Ollila2012}.
    \item \textbf{Residual phase/co-registration:} larger residuals depress background $|\gamma|$, narrowing separation and reducing CCD headroom, underscoring the need for precise co-registration~\cite{Rosen2000,BamlerHartl1998}.
\end{itemize}

It should be noted that the simulator is single-polarisation and single-frequency by default and omits explicit volume/layer and polarimetric effects; nevertheless, it captures first-order links between ($\varepsilon,\sigma,l_c$) and $\{\sigma^0,|\gamma|\}$. Extending to PolSAR would enable Wishart-family change tests and omnibus statistics as likelihood comparators~\cite{Conradsen2003,Akbari2016,Nielsen2020}, and comparing recent SAR-specific deep change detectors would strengthen external validity~\cite{Zhang2023,Cheng2024}.

\section{Conclusion}\label{sec:conclusion}
This paper presented a physics-informed framework for detecting terrain material change in radar imagery, combining an IEM-inspired forward model with a coherence-aware feature stack and unsupervised detectors (RX/robust-RX, CCD, AE). In controlled Monte Carlo experiments spanning non-Gaussian clutter, residual co-registration/phase errors and looks sweep, coherence-centric methods dominated: CCD and a simple score level fusion achieved the highest PR-AUC and F1 at a fixed PFA ($10^{-3}$). Robust scatter (Tyler) stabilised RX in heavy-tailed backgrounds, but RX lagged whenever decorrelation was the principal cue. These findings align with the physics backscatter and decorrelation~\cite{FungChen2004,UlabyLong2014,Touzi1999}, with likelihood-based CCD analysis~\cite{Wahl2016}, and with the role of robust statistics under heavy-tailed radar clutter~\cite{ReedYu1990,Tyler1987,ElDarymli2013}.

\vfill\pagebreak

\bibliographystyle{ieeetr}
\bibliography{main.bib}
\end{document}